\documentclass[runningheads]{llncs}
\usepackage{graphicx}
\usepackage{amsmath}

\usepackage{hyperref}
\usepackage{booktabs}
\usepackage[table]{xcolor}

\begin{document}

\title{Shape Constrained CNN for Cardiac MR Segmentation with Simultaneous Prediction of Shape and Pose Parameters}
\titlerunning{Shape Constrained CNN with Simultaneous Prediction of Shape and Pose}

\author{Sofie Tilborghs \inst{1,4} \and
Tom Dresselaers \inst{2,4} \and
Piet Claus \inst{3,4} \and
Jan Bogaert \inst{2,4} \and
Frederik Maes \inst{1,4}}

\authorrunning{S. Tilborghs et al.}

\institute{Department of Electrical Engineering, ESAT/PSI, KU Leuven, Leuven, Belgium \email{sofie.tilborghs@kuleuven.be} 
\and Department of Imaging and Pathology, Radiology, KU Leuven, Leuven, Belgium \and Department of Cardiovascular Sciences, KU Leuven, Leuven, Belgium \and Medical Imaging Research Center, UZ Leuven, Leuven, Belgium}

\maketitle   

\begin{abstract}
Semantic segmentation using convolutional neural networks (CNNs) is the state-of-the-art for many medical segmentation tasks including left ventricle (LV) segmentation in cardiac MR images. However, a drawback is that these CNNs lack explicit shape constraints, occasionally resulting in unrealistic segmentations. In this paper, we perform LV and myocardial segmentation by regression of pose and shape parameters derived from a statistical shape model. The integrated shape model regularizes predicted segmentations and guarantees realistic shapes. Furthermore, in contrast to semantic segmentation, it allows direct calculation of regional measures such as myocardial thickness. We enforce robustness of shape and pose prediction by simultaneously constructing a segmentation distance map during training. We evaluated the proposed method in a fivefold cross validation on a in-house clinical dataset with 75 subjects containing a total of 1539 delineated short-axis slices covering LV from apex to base, and achieved a correlation of 99$\%$ for LV area, 94$\%$ for myocardial area, 98$\%$ for LV dimensions and 88$\%$ for regional wall thicknesses. The method was additionally validated on the LVQuan18 and LVQuan19 public datasets and achieved state-of-the-art results.   


\end{abstract}

\section{Introduction}
Cardiac magnetic resonance (CMR) imaging provides high quality images of the heart and is therefore frequently used to assess cardiac condition. Clinical measures of interest include left ventricular (LV) volume and myocardial thickness, which can be calculated from a prior segmentation of LV and myocardium. In the last years, convolutional neural networks (CNNs) have shown to outperform traditional model-based segmentation techniques and quickly became the method of choice for this task \cite{Bernard2018}. However, since CNNs are trained to predict a class probability (i.e. LV or background) for each voxel, they are missing explicit shape constraints, occasionally resulting in unrealistic segmentations with missing or disconnected regions and hence requiring postprocessing. In this respect, several authors have proposed to integrate a shape prior in their CNN. Examples are atlases \cite{Duan2019,Zotti2019} or hidden representations of anatomy \cite{Oktay2018,Painchaud2019,Yue2019}. In contrast to CNNs, Active Shape Models (ASM) \cite{Cootes1995} construct a landmark-based statistical shape model from a training dataset and fit this model to a new image using learned local intensity models for each landmark, yielding patient-specific global shape coefficients. In this paper, we combine the advantages of both methods: (1) a CNN is used to extract complex appearance features from the images and (2) shape constraints are imposed by regressing the shape coefficients of the statistical model. Compared to Attar et al. \cite{Attar2019}, who used both CMR images and patient metadata to directly predict the coefficients of a 3D cardiac shape, we enforce robustness of coefficient prediction by simultaneously performing semantic segmentation. A similar approach combining segmentation with regression was used by Vigneault et al. \cite{Vigneault2018} to perform pose estimation of LV, by Gessert and Schlaefer \cite{Gessert2019} and by Tilborghs and Maes \cite{Tilborghs2019} to perform direct quantification of LV parameters and by Cao et al. \cite{Cao} for simultaneous hippocampus segmentation and clinical score regeression from brain MR images. In our approach, the semantic segmentation is performed by regression of signed distance maps, trained using a loss function incorporating both distance and overlap measures. Previous methods to incorporate distance losses include the boundary loss of Kervadec et al. \cite{Kervadec2019}, the Hausdorff distance loss of Karimi and Salcudean \cite{Karimi2020} and the method of Dangi et al. \cite{Dangi} who used separate decoders for the prediction of distance maps and segmentation maps. Different to Dangi et al., our CNN only generates a distance map, while the segmentation map is directly calculated from this distance map, guaranteeing full correspondence between the two representations.

\section{Methods}
\subsection{Shape model}
The myocardium in a short-axis (SA) cross-section is approximated by a set of $N$ endo- and epicardial landmarks radially sampled over uniform angular offsets of $2\pi/N$ rad, relative to an anatomical reference orientation $\theta$. From a training set of images, a statistical shape model representing the mean shape and the modes of variation is calculated using principal component analysis. For each image $i$, the myocardial shape $\textbf{p}_i$ is first normalized by subtracting the LV center position $\textbf{c}_i$ and by rotating around $\theta_i$, resulting in the pose-normalized shape $\textbf{s}_i$:
\begin{equation}
\label{eq:pointnorm}
\begin{bmatrix}
\textbf{s}_{i,x}\\
\textbf{s}_{i,y}
\end{bmatrix}
=
\begin{bmatrix}
\cos(\theta_i) & \sin(\theta_i)\\
-\sin(\theta_i) & \cos(\theta_i)
\end{bmatrix}
\begin{bmatrix}
\textbf{p}_{i,x}-\textbf{c}_{i,x}\\
\textbf{p}_{i,y}-\textbf{c}_{i,y}
\end{bmatrix}
\end{equation}
Representing the shapes as vectors $\textbf{s}_i = (x_{1},...,x_{2N},y_{1},...,y_{2N})$, the mean shape $\overline{\textbf{s}}$ is calculated as $\overline{\textbf{s}} = \frac{1}{I}\sum_{i=1}^{I} \textbf{s}_i$ with $I$ the number of training images. The normalized eigenvectors $\textbf{V} = \{\textbf{v}_1,...,\textbf{v}_m,...,\textbf{v}_{4N}\}$ and corresponding eigenvalues $\lambda_m$ are obtained from the singular value decomposition of the centered shapes $\textbf{s}_i-\overline{\textbf{s}}$. The shape of the myocardium is approximated by the $M$ first eigenmodes:
\begin{equation}
\label{eq:pca}
    \textbf{s}_i \approx \overline{\textbf{s}} + \sum_{m=1}^{M} b_{i,m} \cdot \sqrt{\lambda_m}\cdot \textbf{v}_m
\end{equation}
Using this definition, the variance of the distribution of shape coefficients $b_m$ is the same for every mode $m$.
\subsection{CNN}
\begin{figure*}[tb]
	\centering
	\includegraphics[width =0.99\textwidth]{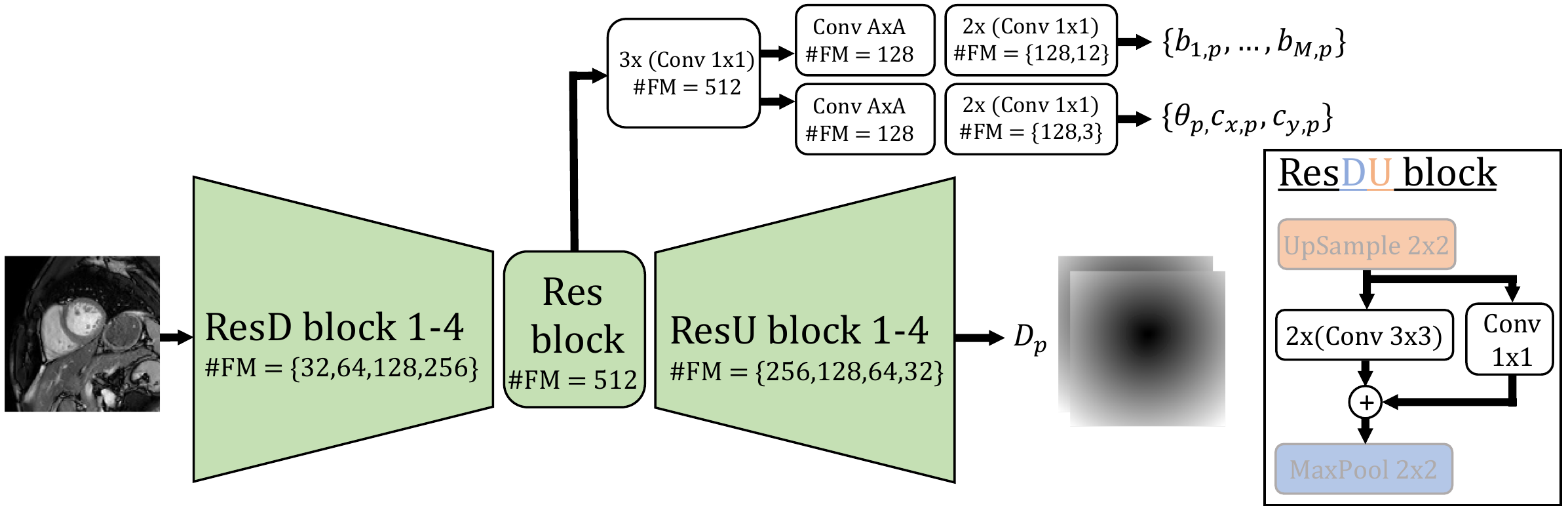}
	\caption{Proposed CNN architecture with three outputs: shape $\{b_{1,p},...,b_{M,p}\}$, pose $\{\theta_p,c_{x,p},c_{y,p}\}$ and distance maps ($D_p$). The details of residual (Res), downsampling Res (ResD) and upsampling Res (ResU) blocks are given on the right. Every convolutional (Conv) layer is followed by batch normalization and a parameterized rectified linear unit, except for the final layer in every output. The number of feature maps ($\#FM$) is the same for all Conv layers in one Res block. The filter size $A$ in a Conv layer is equal to the dimensions of that layer's input. Same padding is used. } 
	\label{fig:network}
\end{figure*}

A schematic representation of the CNN architecture is shown in Fig. \ref{fig:network}. It has three outputs: (1) $M$ predicted shape coefficients $\{b_{1,p},...,b_{M,p}\}$, (2) pose parameters $\{\theta_p,c_{x,p},c_{y,p}\}$ and (3) segmentation map $D_p$. Semantic segmentation is performed by the regression of distance maps $D$. $D$ is an image representing the Euclidean distance $d$ between pixel position and contour. The sign is negative for pixels inside structure $S$:
\begin{equation}
    D(x)=
    \begin{cases}
        -d(x), & \text{if } x \in S\\
        d(x), & \text{if } x \notin S
    \end{cases}
    \label{eq:distmap}
\end{equation}
For both endo- and epicardium, separate distance maps $D_{endo}$ and $D_{epi}$ are created.
The loss function is a weighted sum of the shape loss $L_1$, pose loss $L_2$ and segmentation loss $L_3$:
\begin{equation}
    L = \gamma_1 L_1 + \gamma_2 L_2 + \gamma_3 L_3
\label{eq:totalloss}
\end{equation}
with $L_1$ the mean squared error (MSE) between true and predicted coefficients $b_{m}$: $L_1 = \frac{1}{M} \sum_{m=1}^{M} (b_{m,t}-b_{m,p})^2 $, $L_2$ the MSE for pose parameters $O = \{\theta,c_x,c_y\}$: $L_2 = \frac{1}{3}\sum_{j=1}^3{(o_{j,t}-o_{j,p})^2}$, and $L_3$ a weighted sum of categorical Dice loss and MSE:
\begin{equation}
L_3 = \left(1 - \frac{1}{K} \sum_{k} {\frac{2 \cdot \sum_{x}{S_{k,t}(x)\cdot S_{k,p}(x)}} {\sum_{x}{S_{k,t}(x)} + \sum_{x}S_{k,p}(x)}}\right) +\mu\frac{1}{K\cdot X} \sum_{k,x} (D_{k,t}(x)-D_{k,p}(x))^2
\label{eq:segloss}
\end{equation}
where $X$ is the number of pixels in the image, $K$ is the number of classes and $S_k$ is the binarized distance map using a sigmoid as conversion function:
\begin{equation}
    \label{eq:sigmoidconversion}
    S_k = \frac{e^{-\alpha \cdot D_k}}{1+e^{-\alpha \cdot D_k}}
\end{equation}
where $\alpha$ affects the steepness of the sigmoid function.
\subsection{Implementation details}
Endo- and epicardium are both represented by $N=18$ landmarks and $\theta$ is defined as the orientation of the line connecting the center of LV with the middle of the septum. The network predicts the first $M=$ 12 shape coefficients, representing over 99$\%$ of shape variation. Pose parameters $\theta$, $c_x$ and $c_y$ are normalized to the range [-1,1]. Given the notable difference in magnitude of the different losses, they are weighted with $\gamma_1 = 1$, $\gamma_2 = 10$, $\gamma_3 = 100$ and $\mu = 0.1$. These weights were heuristically defined and assure significant contribution of each of the losses. Parameter $\alpha$ in Eq. \ref{eq:sigmoidconversion} is set to 5 to approximate a binary map with an error of only 6.7e$^{-3}$ for a distance of one pixel from the contour. The network is trained end-to-end over 5000 epochs with Adam optimizer, learning rate 2e-3 and batch size 32.

Online data augmentation is applied by adapting pose and shape parameters. Position and orientation offsets are sampled from uniform distributions between [-40,40]$mm$ and [-$\pi$,$\pi$]rad, respectively. Additionally, shape coefficients were adapted as $b_{m,aug} = b_m + a$, where $a$ is sampled from a uniform distribution between -1 and 1. The input images and distance maps are modified accordingly. For the input image, a thin-plate-spline point-to-point registration is performed using the $2N$ original and augmented landmarks while the distance maps are recreated from the augmented landmarks, connected using cubic spline interpolation, according to Eq. \ref{eq:distmap}. Furthermore, Gaussian noise with standard deviation between 0 and 0.1 is online added to the MR images during training.

\section{Experiments}
The models were constructed and validated in a fivefold cross validation on a clinical dataset ('D1') containing images of 75 subjects (M=51, age = 48.2$\pm$15.6 years) suffering from a wide range of pathologies including hypertrophic cardiomyopathy, dilated cardiomyopathy, myocardial infarction, myocarditis, pericarditis, LV aneurysm... The subjects were scanned on a 1.5T MR scanner (Ingenia, Philips Healthcare, Best, The Netherlands), with a 32-channel phased array receiver coil setup. The endo- and epicardium in end-diastole and end-systole in the SA cine images were manually delineated by a clinical expert. To allow calculation of $\theta$, the RV attachment points were additionally indicated. This resulted in a total of 1539 delineated SA images, covering LV from apex to base. All images of a patient were assigned to the same fold. For each fold, a separate shape model was constructed using the four remaining folds. The images were resampled to a pixel size of 2$mm$x2$mm$ and image size of 128x128, which results in a value of 8 for parameter $A$ in Fig. \ref{fig:network}. 

We validated the performance of our method and the added value of each choice with five different setups: (1) semantic segmentation using categorical Dice loss ('S$_{\mu=0}$'), (2) semantic segmentation using combined loss ('S'): $L = \gamma_3L_3$, (3) regression of shape and pose parameters ('R'): $L = \gamma_1L_1+\gamma_2L_2$, (4) regression and segmentation losses ('RS') as in Eq. \ref{eq:totalloss}, (5) loss as in Eq. \ref{eq:totalloss} and with pose and shape data augmentation ('RS-A$_{ps}$'). For setups 1-4, data augmentation only consisted of the addition of Gaussian noise. Due to faster convergence of training without pose and shape data augmentation, setups 1-4 were only trained for 1000 epochs. For each setup, Dice similarity coefficient (DSC), mean boundary error (MBE) and Hausdorff distance (HD) were calculated from the binarized distance maps ('Map'), as well as from the predicted shape and pose parameters by converting the parameters to landmarks using Eq. \ref{eq:pointnorm} and \ref{eq:pca} and interpolating with cubic splines ('Contour'). The position and orientation errors were respectively defined as $\Delta d = \sqrt{(c_{x,t}-c_{x,p})^2+(c_{y,t}-c_{y,p})^2}$ and $\Delta\theta=|\theta_{t}-\theta_{p}|$. The influence of every shape coefficient was validated by calculating the Euclidean distance between ground truth landmarks and landmarks reconstructed using an increasing number of predicted coefficients. To only capture the impact of shape coefficients, ground truth pose parameters were used for reconstruction. 
Furthermore, LV area, myocardial area, LV dimensions in three different orientations and regional wall thickness (RWT) for six cardiac segments were calculated from the predicted landmarks. LV dimensions and RWT were directly obtained by calculating the distance between two corresponding landmarks and averaging the different values in one segment. For these four physical measures, mean absolute error (MAE) and Pearson correlation coefficient ($\rho$) were calculated. Statistical significant improvement of every choice was assessed by the two-sided Wilcoxon signed rank test with a significance level of 5$\%$.

Additionally, we applied the proposed method to two different public datasets: LVQuan18 \cite{LVQuan18} and LVQuan19 \cite{LVQuan19}. Both datasets contain mid-cavity SA slices for 20 time frames spanning the complete cardiac cycle and provide ground truth values for LV and myocardial area, three LV dimensions and six RWT. In LVQuan18 (145 patients, 2879 images), the 80x80 images were normalized for pose and size while in LVQuan19 (56 patients, 1120 images), no preprocessing was applied. LVQuan19 was identically processed as D1, including prior resampling. Since LVQuan18 contained small, centered images, these images were not resampled, no pose regression was applied, the number of epochs was decreased to 1000 and parameter $A$ in Fig. \ref{fig:network} equals 5. For both datasets, a fivefold cross validation was performed and LV area, myocardial area, LV dimensions and RWT were calculated.

\section{Results}
\begin{figure}[tb]
    \centering
    \includegraphics[width =0.5\linewidth]{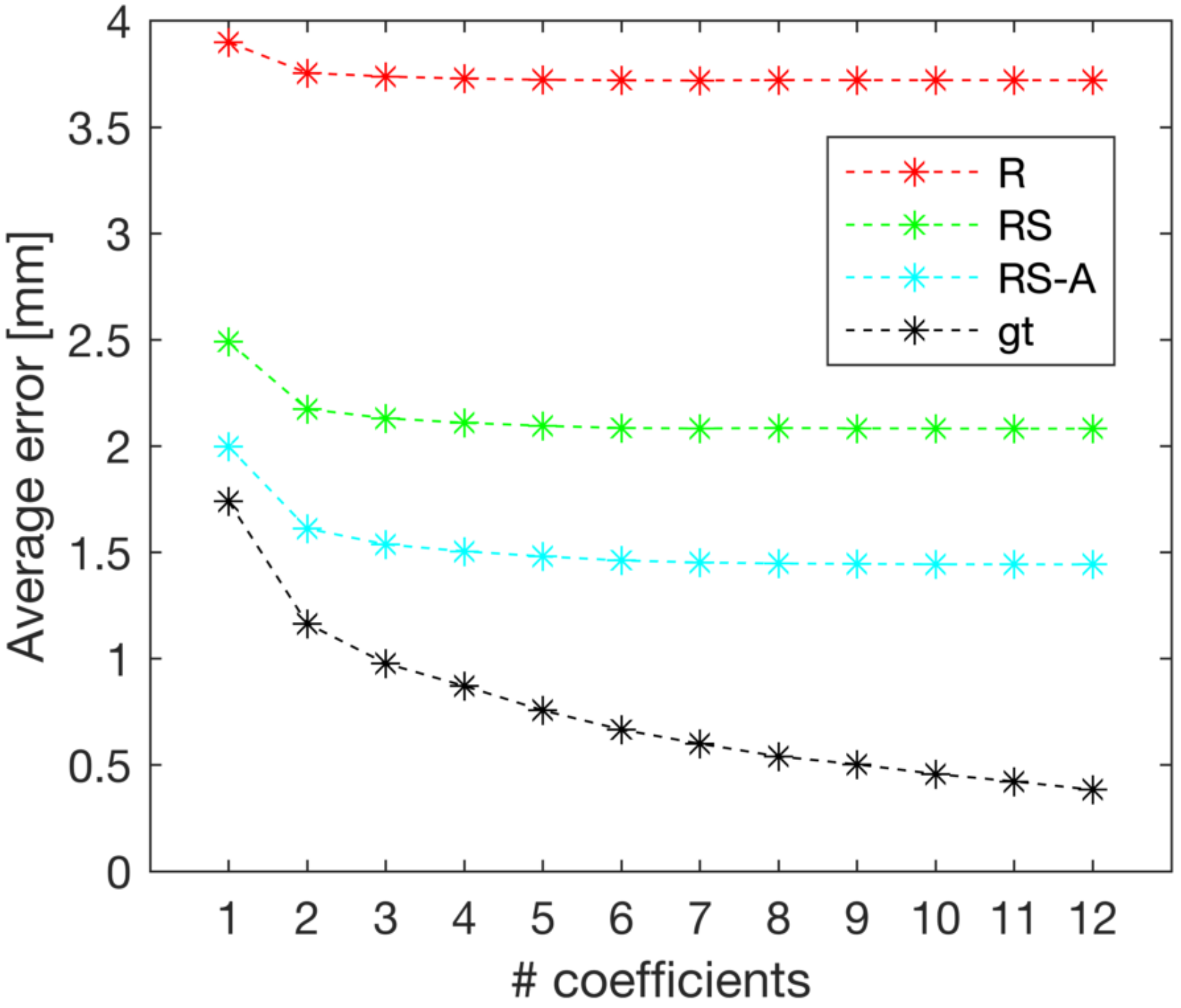}
    \caption{Average distance between ground truth landmarks and landmarks reconstructed using a limited number of coefficients. The results are given for predicted as well as ground truth (gt) coefficients.}
    \label{fig:shapePoints}
\end{figure}

Table \ref{tab:Seg} shows the results of DSC, MBE, HD, $\Delta d$ and $\Delta\theta$ for the different setups. The combined MSE and Dice loss (S) significantly improved DSC, MBE and HD compared to the the setup with only Dice loss (S$_{\mu=0}$), most notably for HD. S$_{\mu=0}$ resulted in 10.2$\%$ unrealistic shapes and S in 0$\%$. While adding $L_1$ and $L_2$ (RS) did not alter the performance of distance map regression, shape and pose data augmentation (RS-A$_{ps}$) did significantly improve all metrics. For the 'Contour' experiments, the addition of semantic segmentation and data augmentation both significantly improved the results, except for $\Delta \theta$. However, DSC, MBE and HD remain worse compared to the 'Map' experiments. The distance errors on the landmarks are visualized in Fig. \ref{fig:shapePoints}, which indicates again that both modifications to a standard regression CNN contribute to significant improvement. Furthermore, whereas the first coefficients, accounting for the largest variation, are relatively well predicted, the latter coefficients were not accurately estimated. The average landmark error for setup RS-A$_{ps}$ using 12 shape coefficients is 1.44$mm$, which is lower than the $MBE$, indicating that the inferior segmentation results are partially due to pose estimation.

\begin{table}[tb]
\centering
	\caption{Results for D1 obtained from the binarized distance maps ('Map') or shape and pose parameters ('Contour'). Mean and standard deviation for DSC, MBE, HD, position error ($\Delta d$) and orientation error ($\Delta \theta$) are reported. Best values are indicated in bold. Statistical significant improvement with respect to the previous row is indicated with $^*$.}
	\label{tab:Seg}
	\begin{tabular}{|l|c|c|c|c|c|c|}
		\hline
		& DSC LV [$\%$] & DSC myo [$\%$] & MBE [$mm$]& HD [$mm$] & $\Delta d$ [$mm$] & $\Delta\theta$ [$^\circ$]\\
		\hline
		\underline{Map}& &&&& &\\
		S$_{\mu=0}$&90.5$\pm$13.9& 81.2$\pm$14.0&1.99$\pm$3.47&18.38$\pm$42.39&/&/ \\
		S &91.7$\pm$12.3$^*$&83.1$\pm$12.6$^*$&1.34$\pm$0.90$^*$&4.32$\pm$6.19$^*$&/&/\\
		RS& 91.8 $\pm$ 11.6& 83.1 $\pm$ 12.4& 1.35$\pm$0.92 & 4.23$\pm$4.29&/&/\\
		RS-A$_{ps}$&\textbf{92.8$\pm$10.1}$^*$&\textbf{85.3$\pm$10.6}$^*$&\textbf{1.18$\pm$0.69}$^*$ & \textbf{3.64$\pm$3.00}$^*$&/&/\\
		\underline{Contour}& &&&&&\\
	    R&65.1$\pm$25.5&38.1$\pm$21.9&7.15$\pm$5.29 & 15.41$\pm$10.70 & 10.1$\pm$9.1&10.4$\pm$10.9\\
		RS&82.6$\pm$18.9$^*$&64.3$\pm$21.5$^*$&3.29$\pm$3.29$^*$&7.70$\pm$7.39$^*$&4.1$\pm$5.4$^*$&11.7$\pm$12.4\\
		RS-A$_{ps}$&\textbf{88.1$\pm$11.9}$^*$&\textbf{72.7$\pm$14.1}$^*$&\textbf{2.16$\pm$1.03}$^*$&\textbf{5.37$\pm$3.49}$^*$&\textbf{2.5$\pm$1.8}$^*$&\textbf{9.5$\pm$7.5}$^*$\\
		\hline
	\end{tabular}
\end{table}

\begin{figure}
    \centering
        \centering
        \begin{minipage}[t]{0.3\linewidth}
    		\includegraphics[width = \linewidth]{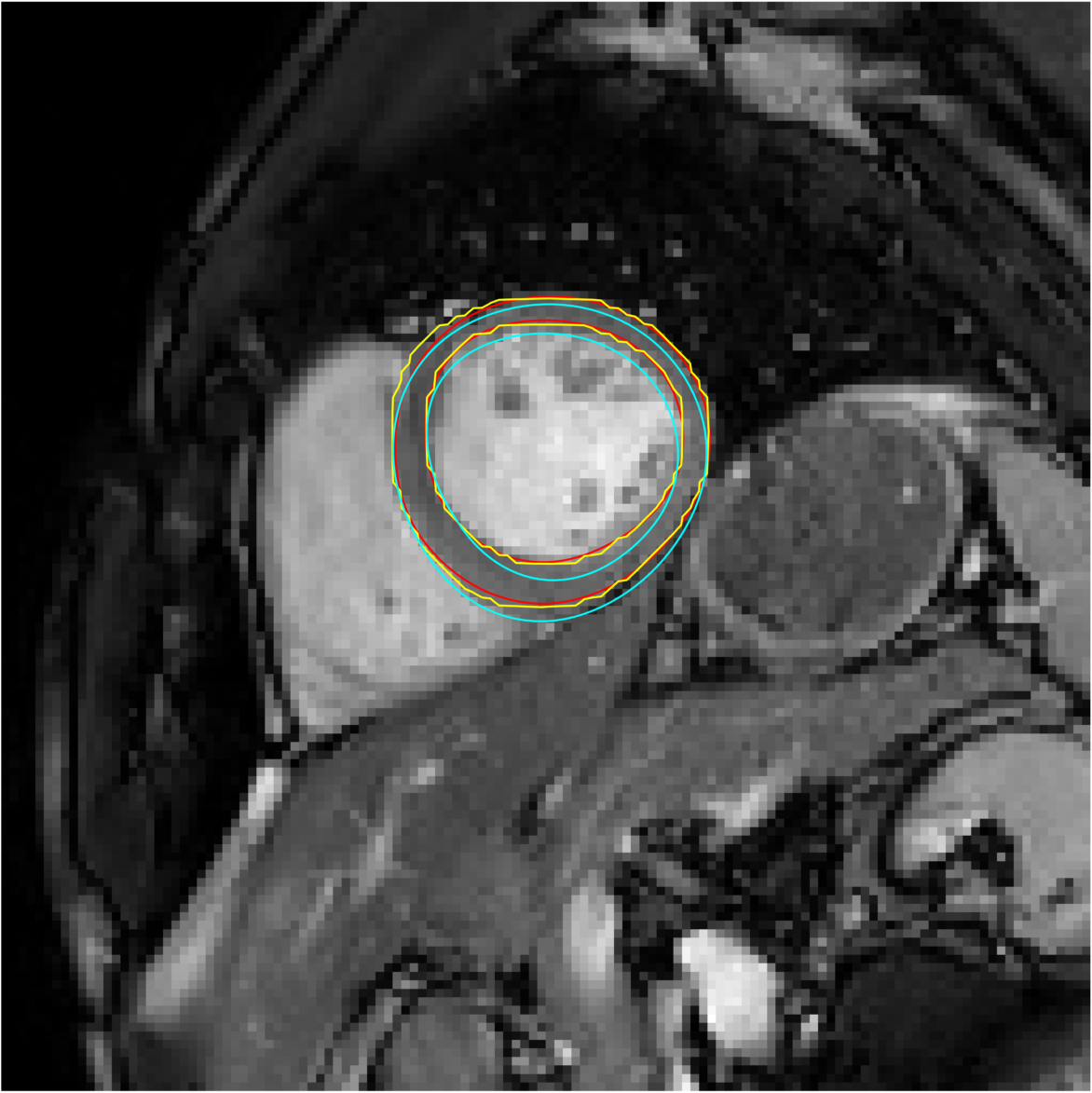}
	    \end{minipage} 
        \begin{minipage}[t]{0.3\linewidth}
    		\includegraphics[width = \linewidth]{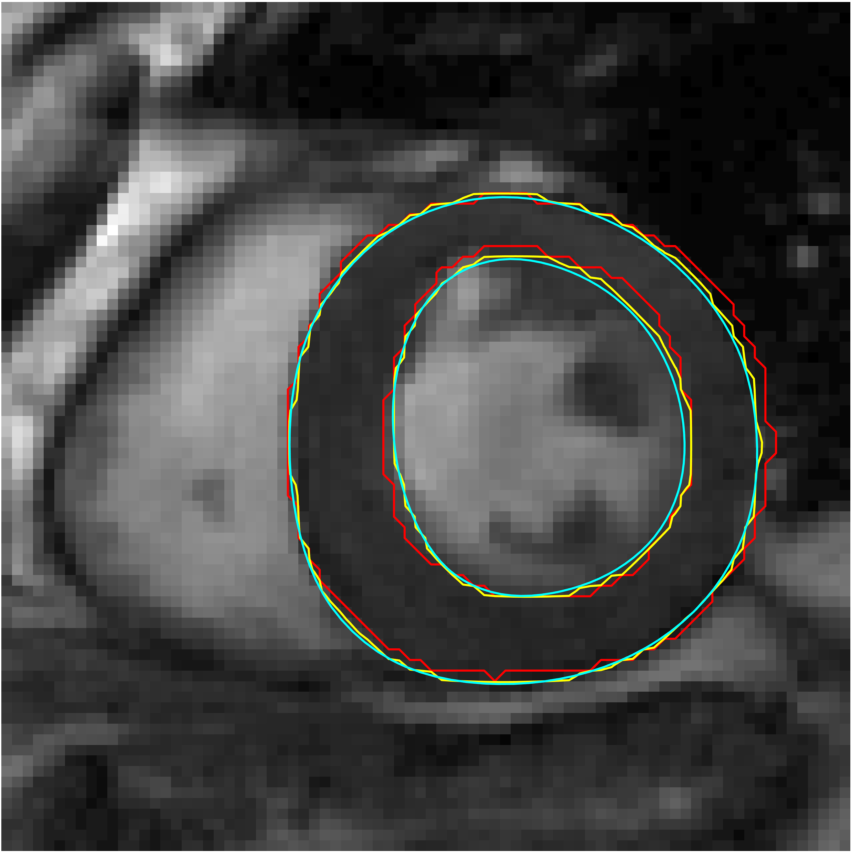}
	    \end{minipage} 
	    \begin{minipage}[t]{0.3\linewidth}
    		\includegraphics[width = \linewidth]{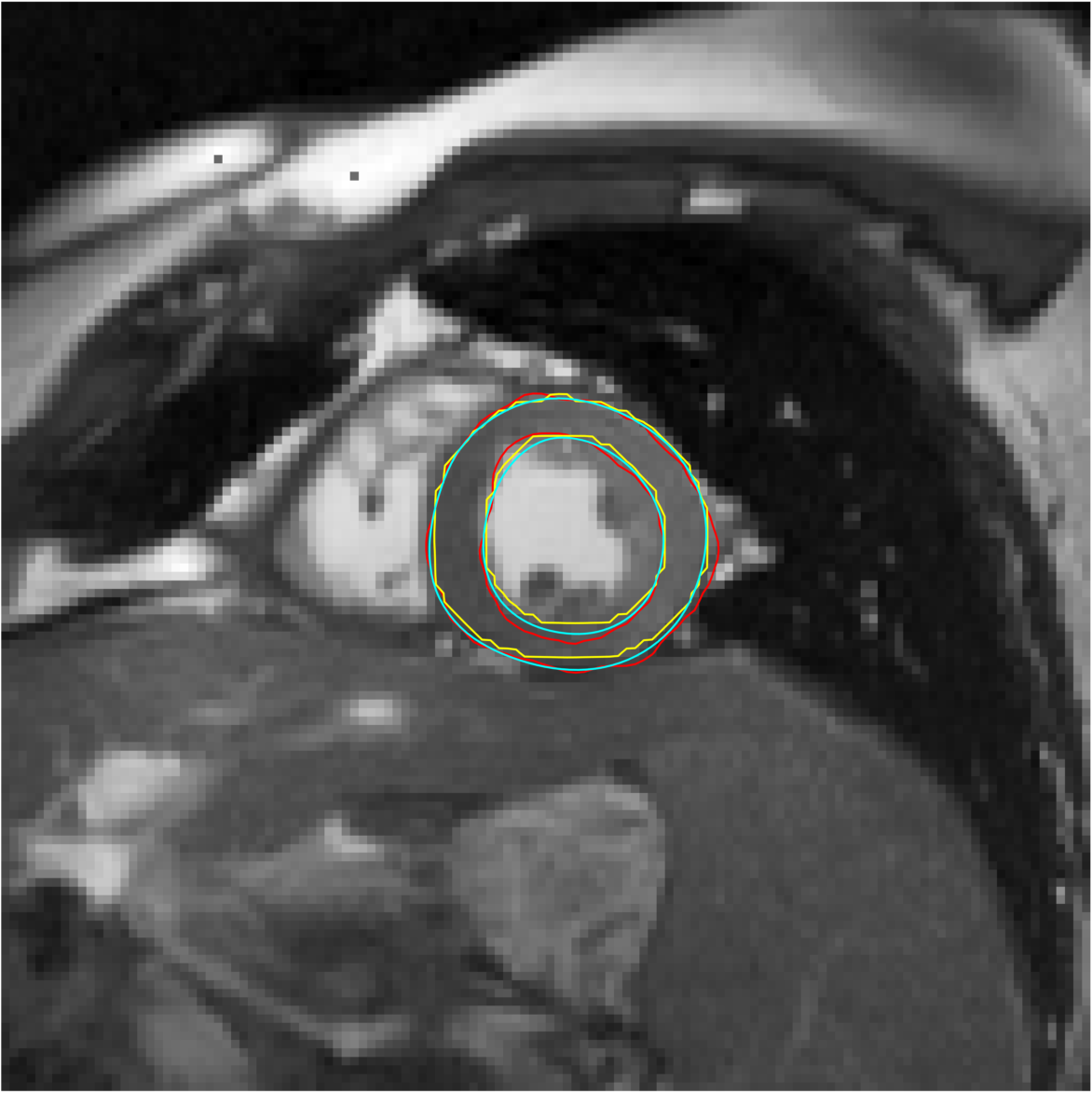}
	    \end{minipage} 
    
        \begin{minipage}[t]{0.3\linewidth}
    		\includegraphics[width = \linewidth]{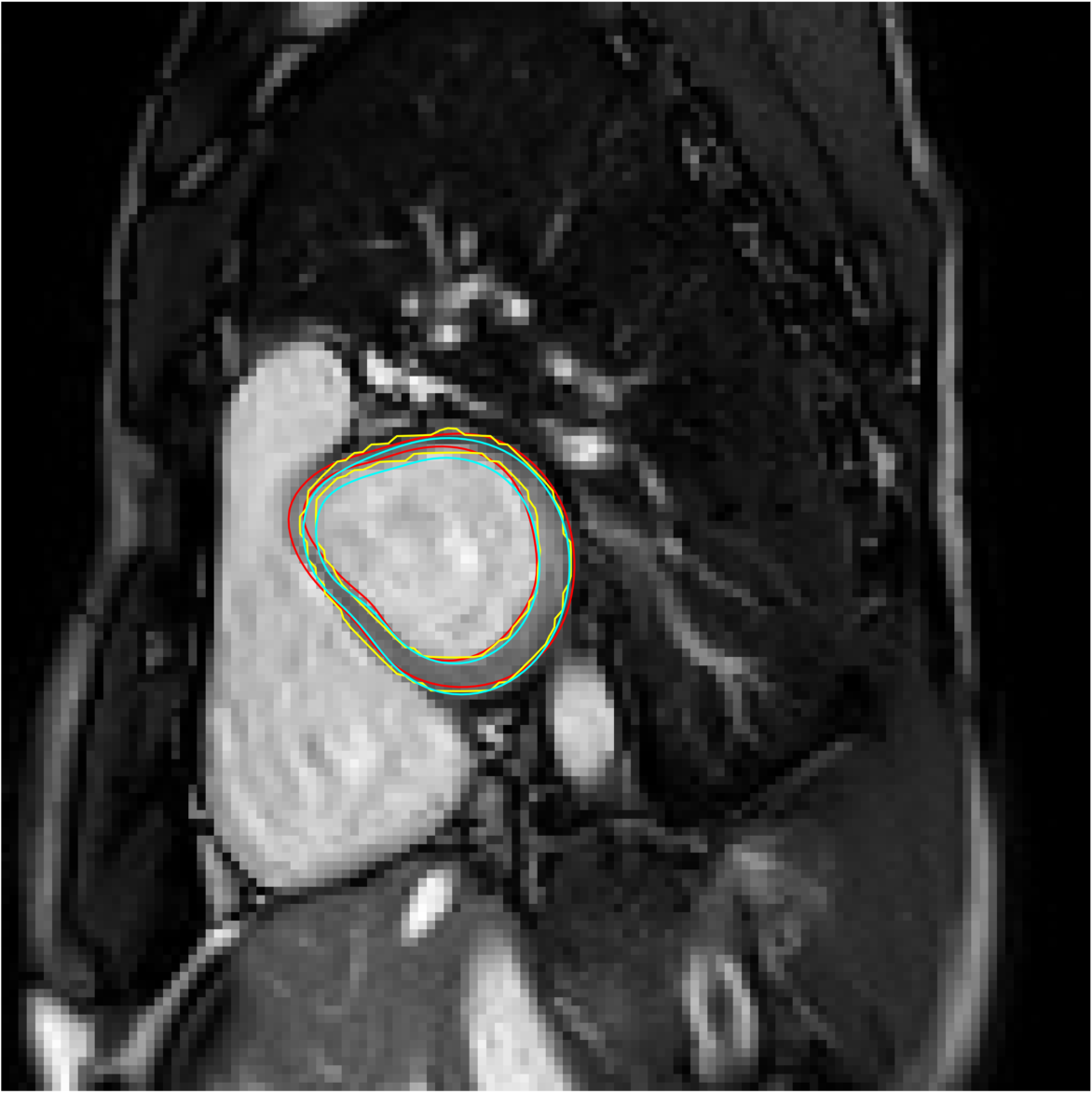}
	    \end{minipage} 
	    \begin{minipage}[t]{0.3\linewidth}
    		\includegraphics[width = \linewidth]{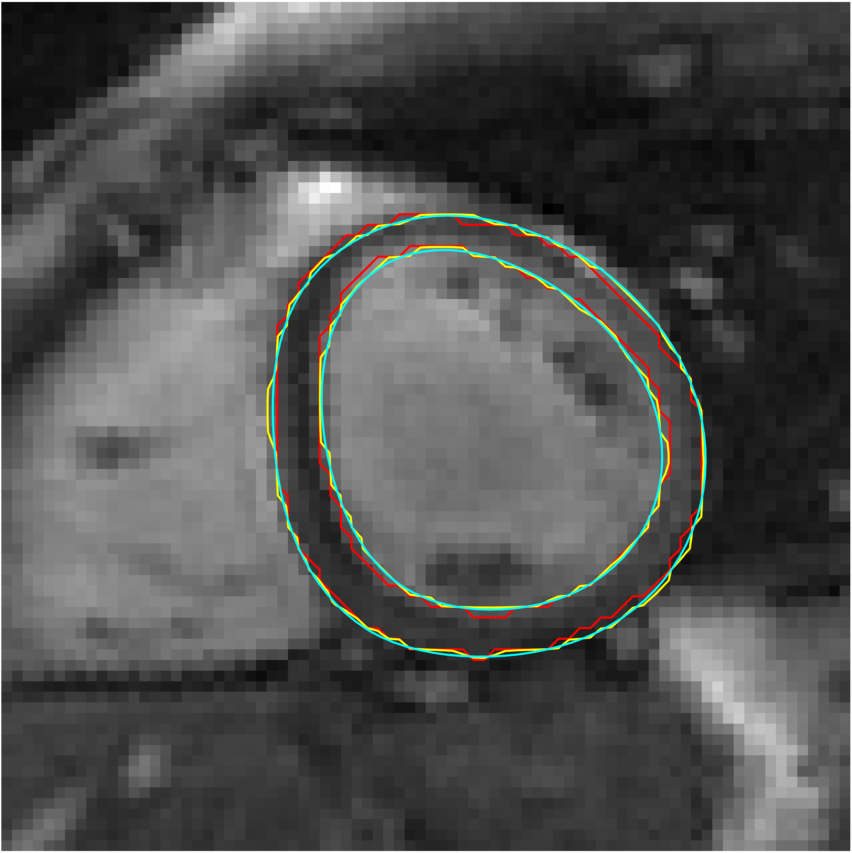}
	    \end{minipage} 
	    \begin{minipage}[t]{0.3\linewidth}
    		\includegraphics[width = \linewidth]{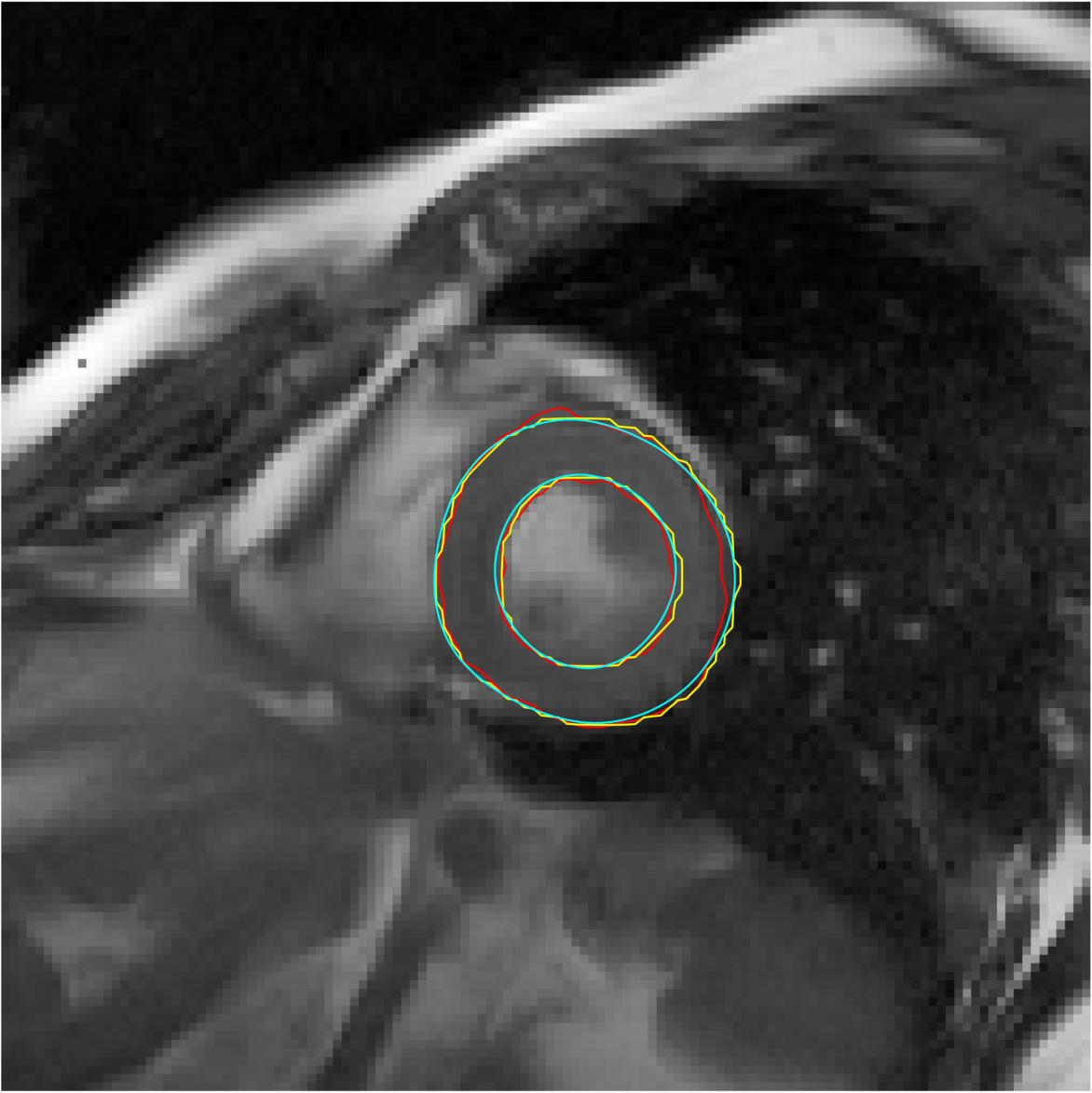}
	    \end{minipage} 
        \caption{Representative segmentation examples for datasets D1, LVQuan18 and LVQuan19 (left to right). Ground truth (red), semantic segmentation output (yellow) and segmentation reconstructed from shape and pose output (cyan) are shown.}
        \label{fig:Seg}
\end{figure}

Table \ref{tab:physical} reports MAE and $\rho$ of LV area, myocardial area, LV dimensions and RWT, averaged over all segments. The results on D1 show that these metrics can be more accurately estimated by simultaneous semantic segmentation and by addition of data augmentation. For LV and myocardial area and LV dimensions, RS-A$_{ps}$ obtained better results compared to the winner of the LVQuan18 challenge \cite{winnerLVquan18}, who used a parameter regression approach, while the estimation of RWT was slightly worse. For LVQuan19, the results of RS-A$_{ps}$ are compared to the top three entries of the challenge. While the results of \cite{winnerLVquan19} and \cite{Gessert2019} are superior, our error on LV and myocardial area and LV dimensions is lower compared to the errors reported in \cite{Tilborghs2019}, and the correlation is higher for all metrics.
Fig. \ref{fig:Seg} depicts representative segmentation examples.
\begin{table}[tb]
    \centering
	\caption{MAE and $\rho$ for LV area, myocardial area, LV dimensions and RWT. Best values are indicated in bold. For D1, statistical significant improvement with respect to the previous column is indicated with $^*$. $^{(1)}$In \cite{winnerLVquan19}, a threefold cross validation was used. $^{(2)}$In \cite{Gessert2019}, the average MAE of LV and myocardial area was reported to be 122$mm^2$.}.
	\label{tab:physical}
	\begin{tabular}{|l|l|c|c|c||c|c||c|c|c|c|}
		\hline
		\multicolumn{2}{|c|}{}&\multicolumn{3}{c||}{D1 [$\%$]} & \multicolumn{2}{c||}{LVQuan18}& \multicolumn{4}{c|}{LVQuan19}\\
	    \multicolumn{2}{|c|}{}& R & RS & RS-A$_{ps}$ &  \cite{winnerLVquan18} & RS-A$_{ps}$  & \cite{winnerLVquan19}$^1$& \cite{Gessert2019} & \cite{Tilborghs2019} & RS-A$_{ps}$\\
		\hline
		MAE&Area LV [$mm^2$] & 472 & 256$^*$ &\textbf{139}$^*$ & 135&\textbf{117} & \textbf{92} &122$^2$&186& 134\\
		&Area Myo [$mm^2$]  & 299 & 192$^*$ & \textbf{145}$^*$ &177&\textbf{162} & \textbf{121} &122$^2$&222& 201\\
		&Dim [$mm$] & 7.06 & 3.58$^*$& \textbf{2.37}$^*$& 2.03&\textbf{1.50} & \textbf{1.52} &1.84&3.03&2.10\\
		&RWT [$mm$]& 1.86 & 1.38$^*$ & \textbf{1.18}$^*$&\textbf{1.38}&1.52 &\textbf{1.01} &1.22&1.67&1.78\\
		\hline
		$\rho$ $[\%]$&Area LV& 81 & 95 &\textbf{99} & / & 99 & / &/&97& \textbf{98} \\
		&Area Myo& 77 & 90 & \textbf{94} & / & 93 & / &/&88& \textbf{93} \\
		&Dim& 84 & 96 & \textbf{98} & / & 98 & / &/&95& \textbf{97} \\
		&RWT& 69 & 83 & \textbf{88} & / & 84 & / &/&73& \textbf{83} \\
		\hline
	\end{tabular}
\end{table}

\section{Discussion}
In contrast to semantic segmentation, the predicted shape coefficients are directly linked to an oriented landmark-based representation and as such allow straightforward calculation of regional metrics including myocardial thickness or strain. Furthermore, contrary to conventional semantic segmentation using Dice loss (S$_{\mu=0}$), our approach did not result in any missing or disconnected regions since the shape model is inherently unable to predict such unrealistic shapes. While some initial experiments showed that pose and shape data augmentation was able to significantly improve the segmentation for setup S$_{\mu=0}$, the results remained significantly worse compared to the proposed approach RS-A$_{ps}$.

For the LVQuan19 challenge data, we obtained higher MAE compared to the leading results of \cite{winnerLVquan19}.
There are multiple possible explanations for this. First, the two methods use significantly different approaches: Acero et al. \cite{winnerLVquan19} calculated the LV parameters from a segmentation obtained with a semantic segmentation CNN while we calculated the LV parameters from the 12 predicted shape coefficients. When calculating the LV parameters from the predicted distance maps and position instead, slightly lower MAE of 109$mm^2$ for LV area, 188$mm^2$ for myocardial area, 1.69$mm$ for LV dimensions and 1.74$mm$ for RWT were achieved. This is in accordance with the lower performance of the 'Contour' experiments compared to the 'Map' experiments in Table 1. Second, preprocessing steps such as resampling strategy and intensity windowing, data augmentation and training approach all have an impact on CNN performance. In the LVQuan18 challenge, the images were preprocessed by the challenge organizers, eliminating some of these sources of variability.
Third, contrary to the challenge entries \cite{winnerLVquan19,Gessert2019,Tilborghs2019}, our method was not specifically developed and tuned for this task. It should be noted that all three challenge entries reported substantially worse results on LVQuan19 test set, which is not publicly available.

We found that the regression of shape coefficients is a more difficult task compared to semantic segmentation. In semantic segmentation using distance maps, 128x128 correlated values should be estimated for every image while shape coefficient regression required the estimation of 12 uncorrelated values from relatively little training data. The combination with semantic segmentation and addition of data augmentation was however able to significantly improve the shape coefficient regression. In future work, we want to investigate if an extra loss term enforcing consistency between semantic segmentation and pose and shape parameters can further improve these results. 

\section{Conclusion}
In this paper, we presented a proof-of-concept of our shape constrained CNN on 2D cardiac MR images for segmentation of LV cavity and myocardium. In the future, this can be expanded to 3D segmentation and to other applications.

\subsubsection{Acknowledgement}
Sofie Tilborghs is supported by a Ph.D fellowship of the Research Foundation - Flanders (FWO). The computational resources and services used in this work were provided in part by the VSC (Flemisch Supercomputer Center), funded by the Research Foundation - Flanders (FWO) and the Flemisch Government - department EWI. This research also received funding from the Flemish Government under the “Onderzoeksprogramma Artificiële intelligentie (AI) Vlaanderen" programme and is also partially funded by KU Leuven Internal Funds C24/19/047 (promotor F. Maes).

\end{document}